\documentclass[10pt, journal]{IEEEtran} 

\usepackage{graphicx,amssymb,amsmath}
\usepackage{multicol}
\usepackage[noadjust]{cite}
\usepackage{setspace}
\usepackage{subfigure}
\usepackage{graphicx}
\usepackage{float}
\usepackage{url}
\usepackage{stfloats}
\usepackage{amsthm,pifont}
\usepackage{flushend}
\usepackage{cases,subeqnarray}
\usepackage{bm,multirow,bigstrut}
\usepackage{amsmath, amsthm, amssymb}
\usepackage{textcomp}
\usepackage{latexsym,bm}
\usepackage{booktabs}
\usepackage{mathtools}
\usepackage{dsfont}
\usepackage{extarrows}
\usepackage{epsfig}
\usepackage{epsfig}
\usepackage{epstopdf}
\usepackage[noend]{algpseudocode}
\usepackage{makecell}
\usepackage[linesnumbered,ruled]{algorithm2e}
\usepackage[table]{xcolor}
\theoremstyle{plain}

\theoremstyle{plain}

\usepackage{amsmath}
\newtheorem{theorem}{Theorem}

\IEEEoverridecommandlockouts
\allowdisplaybreaks[4]  

\begin{document}
	\title{AOLO: Analysis and Optimization For Low-Carbon Oriented Wireless Large Language Model Services}
	\author{Xiaoqi Wang, Hongyang Du, Yuehong Gao, and Dong In Kim,~\IEEEmembership{Life Fellow, IEEE}
		\thanks{X. Wang and Y. Gao are with the School of Information and Communication Engineering, Beijing University of Posts and Telecommunications, Beijing, China (e-mail: xiaoqi\_wang@bupt.edu.cn, yhgao@bupt.edu.cn).}
		\thanks{H. Du is with the Department of Electrical and Electronic Engineering, University of Hong Kong, Pok Fu Lam, Hong Kong (e-mail: duhy@eee.hku.hk).}
		\thanks{D. I. Kim is with the College of Information and Communication Engineering, Sungkyunkwan University, South Korea (e-mail: dongin@skku.edu).}
	}
	\maketitle
	\begin{abstract}
Recent advancements in large language models (LLMs) have led to their widespread adoption and large-scale deployment across various domains.
However, their environmental impact, particularly during inference, has become a growing concern due to their substantial energy consumption and carbon footprint.   
Existing research has focused on inference computation alone, overlooking the analysis and optimization of carbon footprint in network-aided LLM service systems. 
To address this gap, we propose AOLO, a framework for {\underline{a}}nalysis and  {\underline{o}}ptimization for  {\underline{l}}ow-carbon  {\underline{o}}riented wireless LLM services. AOLO introduces a comprehensive carbon footprint model that quantifies greenhouse gas emissions across the entire LLM service chain, including computational inference and wireless communication.
Furthermore, we formulate an optimization problem aimed at minimizing the overall carbon footprint, which is solved through joint optimization of inference outputs and transmit power under quality-of-experience and system performance constraints.
To achieve this joint optimization, we leverage the energy efficiency of spiking neural networks (SNNs) by adopting SNN as the actor network and propose a low-carbon-oriented optimization algorithm, i.e., SNN-based deep reinforcement learning (SDRL).
Comprehensive simulations demonstrate that SDRL algorithm significantly reduces overall carbon footprint, achieving an 18.77\% reduction compared to the benchmark soft actor-critic, highlighting its potential for enabling more sustainable LLM inference services.

	\end{abstract}
	\IEEEpeerreviewmaketitle

	\section{Introduction}

Large language models (LLMs) and generative artificial intelligence (GenAI) technologies have experienced rapid advancements, showing remarkable capabilities in generating text, images, videos, and multimodal content, enabling diverse applications across creative, informational, and commercial domains \cite{chang2024survey,cao2024survey}. 
The rapid industrial adoption of LLMs is exemplified by ChatGPT, which reached 100 million monthly active users within just two months of its launch and continues its growth, now attracting 464 million monthly visitors as of November 2024 \cite{Duarte2025}.
Building on this success, global technology giants, such as Google, Microsoft and Meta, have launched a series of applications powered by LLM technology, accelerating its  adoption across diverse sectors \cite{touvron2023llama,team2023gemini}.

The transformative impact of LLMs on numerous industries is undeniable, yet their rapid growth raises mounting concerns regarding sustainability and environmental impacts \cite{luccioni2023estimating,jiang2024preventing}.
Both the training and inference phases of LLMs demand immense computational resources, resulting in considerable energy consumption and carbon footprint.
According to a report by the Stanford Institute for Human-Centered Artificial Intelligence \cite{maslej2023artificial}, training OpenAI's GPT-3 requires 1287 megawatt-hours (MWh) of electricity, which is comparable to the energy consumed by an average American household over 122 years\footnote{According to the U.S. Energy Information Administration, the average residential utility customer consumes approximately 10500 kilowatt-hours (kWh) of electricity annually \cite{EIA_Electricity_Use_in_Homes}.}, and results in an estimated 552 tonnes of CO2 equivalent emissions.
Such staggering energy consumption and CO2 emissions underscore the environmental challenges of large-scale GenAI models, especially LLMs, and raise urgent concerns about long-term sustainability as their scale and complexity continue to grow.

While environmental costs of the LLM training  phase  are widely recognized \cite{li2024towards,liu2024green,wu2022sustainable}, the inference phase, due to its continuous and large-scale deployment, has emerged as an even larger contributor to the carbon footprint. 
For example, ChatGPT processes over 270 million daily inference requests \cite{chien2023reducing}, and its inference-related carbon footprint matches its training footprint in just 121 days \cite{fu2024llmco2}.
With the growing adoption of LLM applications across diverse sectors, 
the increasing volume of daily inference requests accelerates this pattern - inference emissions now surpass training emissions in shorter timeframes, emphasizing the importance of actively monitoring and mitigating the carbon footprint associated with the inference phase. 
Moreover, recent research on scaling inference-time computation has demonstrated that increasing computational resources during inference can enhance model performance \cite{brown2024large,snell2024scaling}, as evidenced by the recent success of DeepSeek-V3 \cite{liu2024deepseek}.
In the future, more computational resources will be required for  LLM inference to achieve better model performance, which will inevitably lead to higher energy consumption and an increased carbon footprint associated with LLM inference.

In response, researchers have increasingly prioritized the development of tools and strategies to better quantify and mitigate its environmental impact.
On the one hand, several tools and methodologies have been introduced to enhance the precision of carbon footprint estimations for LLM inference \cite{faiz2024llmcarbon,fu2024llmco2,budennyy2022eco2ai}.
LLMCarbon, proposed in \cite{faiz2024llmcarbon}, offers an end-to-end carbon footprint model that captures both operational and embodied emissions of dense and sparse LLMs, providing a comprehensive  estimation framework.
LLMCO2 \cite{fu2024llmco2} employs a graph neural network-based approach to significantly improve the prediction accuracy of inference-related carbon footprint by incorporating hardware-specific features.
Eco2AI \cite{budennyy2022eco2ai}, a real-time tracking tool, integrates hardware metrics and regional electricity grid data to monitor energy consumption and CO2 emissions during inference.
On the other hand, carbon-aware optimization strategies have been explored to reduce the carbon footprint of LLM inference. 
For instance, SPROUT, introduced in \cite{li2024toward}, reduces emissions by employing token generation directives, enabling carbon reductions while maintaining high-quality outputs.
Similarly, CarbonMin, proposed in \cite{chien2023reducing}, is a carbon-aware strategy designed to dynamically allocate inference workloads to regions with lower grid carbon intensity, achieving up to 56\% emission reductions in real-world scenarios.

While these approaches estimate and optimize the carbon footprint of inference computation, they overlook a critical component: the wireless transmission of inference outputs from data centers to users. This transmission phase contributes significantly to the overall carbon footprint \cite{li2023carbon,ding2022carbon}, involving energy-intensive operations such as signal modulation, channel coding, and radio frequency transmission.
The growing size of LLM inference outputs and stringent low-latency transmission requirements further highlight the increasing importance of considering the emissions at the wireless transmission phase for providing LLM inference service to users.
Hence, a comprehensive carbon footprint estimation model that integrates emissions from both inference computation and wireless communication is essential for accurately assessing and effectively mitigating the overall carbon impact of LLM inference service systems.

Considering the complex dynamics of wireless transmission and the variability of inference requirements, deep reinforcement learning (DRL) is an effective approach for optimizing the overall carbon footprint of LLM inference due to its adaptability to dynamic environments.
However, while DRL has achieved remarkable success across diverse domains, its reliance on energy-intensive multilayer perceptrons (MLPs) poses significant challenges to sustainability \cite{anthony2020carbontracker,yamazaki2022spiking}. 
The integration of spiking neural networks (SNNs) into DRL frameworks \cite{yuan2019reinforcement,tang2021deep,zhang2022multi}, particularly through adopting SNNs as actor networks, presents a promising solution. Leveraging event-driven computation and sparse data handling \cite{davies2018loihi}, SNNs can significantly reduce energy consumption while maintaining robust performance \cite{wang2023event,naya2021spiking}.
This makes SNN-based DRL frameworks well-suited for optimizing carbon footprint.

Motivated by the issues mentioned above, we conduct a comprehensive study to estimate and mitigate the carbon footprint of LLM inference services, focusing on both inference computation and wireless communication phases. 
The contributions of this paper are summarized as follows:
	\begin{itemize}
	\item We propose AOLO, a framework for {\underline{a}}nalysis and  {\underline{o}}ptimization for  {\underline{l}}ow-carbon  {\underline{o}}riented wireless LLM services, which introduces a comprehensive carbon footprint model that quantifies the overall greenhouse gas emissions associated with serving one LLM inference request. 
	To the best of our knowledge, this is the first model that incorporates the carbon footprint from both the inference computation phase and the wireless communication phase, enabling joint optimization. 	
	
	\item To provide more sustainable LLM inference services, we formulate an optimization problem aimed at minimizing the overall carbon footprint while maintaining user's subjective quality of experience (QoE) and system performance. 
	Considering the inherent remarkable energy efficiency of SNNs, which aligns seamlessly with the goal of reducing environmental impact, we propose an SNN-based DRL (SDRL) algorithm to achieve low-carbon-oriented optimization, which generates optimal decisions under dynamic environmental conditions to optimize the overall carbon footprint.
	
	\item We validate the effectiveness of the proposed SDRL algorithm through extensive simulations and comparisons with benchmark solutions. Additionally, we analyze the influence of key hyperparameters in the SNN to provide deeper insights into the efficiency and behavior of SDRL.
	\end{itemize}
	
	The remainder of this paper is organized as follows: 
	Section \ref{SEC_RelatedWork} reviews the related work. 
	Section \ref{sec:System Model} introduces the wireless network-aided LLM inference service system and the overall carbon footprint model. 
	Section \ref{SEC_carbon_optimize} details the overall carbon footprint model in terms of inference computation and wireless communication phases, and formulates the  carbon footprint optimization problem. 
	In Section \ref{SEC_SDRL}, we introduce the preliminaries of adopting SNN and propose the SDRL algorithm.
	A comprehensive performance evaluation is conducted in Section \ref{sec:evaluation}, and Section \ref{sec:conclusion} concludes this paper.

	\section{Related Work}\label{SEC_RelatedWork}
	In this section, we provide a brief review of the related work, including LLM services, carbon footprint, DRL and SNN.
	
	\subsection{LLM Services}
LLMs have emerged as a transformative technology in natural language processing, leveraging vast amounts of text data to learn complex language patterns.
With rapid advancements in model architectures and training techniques, these models have achieved unprecedented fluency and accuracy in tasks such as natural language understanding and text generation.
The continuous evolution of LLMs has made them a cornerstone in GenAI-driven applications, enabling breakthroughs across multiple domains. Models, such as GPT \cite{achiam2023gpt}, LLaMA \cite{touvron2023llama} and BERT \cite{devlin2018bert}, have garnered significant attention due to their exceptional comprehension and reasoning capabilities, driving progress in diverse fields. Their outstanding ability to process and generate human-like text has facilitated advancements in areas such as healthcare \cite{singhal2025toward}, finance \cite{wu2023bloomberggpt}, education \cite{yan2024practical}, and law \cite{lai2024large}. In these domains, LLMs are employed for tasks such as text generation, translation, sentiment analysis, and question answering, significantly enhancing efficiency and decision-making processes.
	\subsection{Carbon Footprint}
Global warming, driven by increasing levels of greenhouse gases, is one of the most critical challenges for humanity. The carbon footprint, which measures the total greenhouse gas emissions linked to a product or service, has become a key metric for assessing the environmental impact of human activities.
The rapid growth and widespread adoption of LLMs have raised concerns about their environmental impact, as the immense computational resources required for LLM inference significantly contribute to the corresponding carbon footprint.
Current research on the carbon footprint of LLM inference primarily focuses on two main aspects: 
i) the exploration of tools and methodologies to improve the accuracy of carbon footprint estimation for LLM inference, such as LLMCarbon \cite{faiz2024llmcarbon}, LLMCO2 \cite{fu2024llmco2}, and Eco2AI \cite{budennyy2022eco2ai};
ii) the exploration of carbon-aware optimization strategies aimed at reducing the carbon footprint of LLM inference, including SPROUT \cite{li2024toward}, which reduces emissions through token generation directives, and CarbonMin \cite{chien2023reducing}, which dynamically allocates inference workloads to regions with lower grid carbon intensity.
However, these studies have overlooked the carbon footprint associated with the transmission of inference results, an important aspect of the overall carbon footprint. 
Several studies have assessed the carbon footprint of transmission systems, such as 5G base stations and mobile networks \cite{li2023carbon,ding2022carbon}.
There remains a gap in research concerning the comprehensive assessment and optimization of the carbon footprint across the entire LLM inference service, including both inference computation and wireless communication phases.

	\subsection{DRL and SNN}
DRL has gained widespread attention as an effective approach for addressing high-dimensional and dynamic challenges  across diverse domains, ranging from robotic control tasks \cite{tang2024deep,xie2023drl,calderon2024deep} to complex network optimization problems \cite{fang2023drl,zhang2023security,zhang2023drl}. 
Within the DRL framework, MLPs, a type of fully connected deep neural networks (DNNs), have been widely adopted as the backbone for the actor and critic networks. 
This combination has enabled DRL to achieve notable success in tackling diverse real-world challenges. 
However, the dense matrix operations and continuous updates required by DNNs, particularly MLPs, make them highly resource-intensive in computation and energy, posing significant drawbacks \cite{yamazaki2022spiking}. 
Consequently, MLP-based DRL frameworks incur substantial energy consumption, especially in scenarios involving continuous interactions with dynamic environments. Such inefficiencies hinder the long-term sustainability and scalability of DRL systems, particularly in energy-constrained settings.

SNNs offer a promising alternative to traditional DNNs by addressing the critical issue of energy efficiency. 
Drawing inspiration from biological neural systems, SNNs utilize event-driven computation to process information through discrete spikes, enabling more efficient and sparse data handling \cite{davies2018loihi}. 
This mechanism results in significantly reduced energy consumption while maintaining competitive performance across a range of tasks \cite{strohmer2020flexible,wang2023event,naya2021spiking}. As a result, SNNs have emerged as a desirable  option for low-carbon and energy-sensitive applications. 
Building on the advantages of SNN, there is growing interest in integrating SNN into DRL frameworks \cite{yuan2019reinforcement,tang2021deep,zhang2022multi}, offering a promising solution to the energy inefficiencies inherent in traditional MLP-based DRL architectures.

Considering the existing research limitations, we conduct a comprehensive study to estimate and mitigate the carbon footprint of LLM inference services, focusing on both the inference computation and wireless communication phases. To this end, we formulate an optimization problem aimed at minimizing the overall carbon footprint and propose the SDRL algorithm to achieve low-carbon-oriented optimization.

	\section{System Model}\label{sec:System Model}
	
	In this section, we present the wireless network-aided LLM service system and introduce the overall carbon footprint model associated with serving one LLM inference request. 
	
	\subsection{Wireless Network-Aided LLM Service System}	
\begin{figure}[t]
		\centering
		\includegraphics[width=0.49\textwidth]{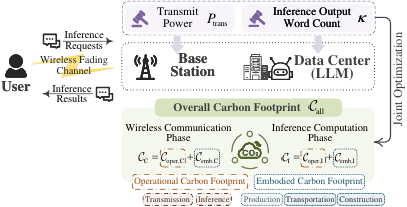}
		\caption{Illustration of wireless network-aided LLM inference service system with a data center and a base station. 
		The overall carbon footprint is minimized through joint optimization of inference outputs and transmit power.
		}
		\label{SystemModel}
	\end{figure}

	We consider a network-aided LLM inference service system, where the inference requests are processed by a data center, and the inference outputs are transmitted to users via a wireless communication network facilitated by a base station (BS), as illustrated in Fig. \ref{SystemModel}. In this system, users send inference requests to the data center, where computational resources are allocated to generate the desired inference  outputs. These outputs are subsequently delivered to the users through wireless transmission enabled by the BS.
	
	Nakagami-$ m $ fading distribution is adopted to model the wireless channel between the user and server \cite{nakagami1960m}. 
	For the further performance analysis, the investigation of the instantaneous signal-to-noise ratio (SNR), denoted by $ \gamma  = \frac{P_{\text{trans}}X}{{{\sigma ^2}}} $, is essential, where $ X={\left| h \right|^2} $ represents the channel power gain, $ \left| h \right| $ is the fading envelope, $ P_{\text{trans}} $ is the transmit power, and $ \sigma ^2 $ is the noise variance.
	With the help of the probability density function (PDF) and cumulative distribution function (CDF) of Nakagami-$ m $ fading  envelope $ \left| h \right| $ given by \cite{nakagami1960m}, the corresponding PDF and CDF of the instantaneous SNR can be expressed as
	\begin{align}\label{pdf_snr}
		f\left( \gamma  \right) = \frac{{{m^m}{\sigma ^{2m}}}}{{\Gamma \left( m \right){\Omega ^m}P_{{\rm{trans}}}^m}}{\gamma ^{m - 1}}\exp \left( { - \frac{{m{\sigma ^2}}}{{\Omega {P_{{\rm{trans}}}}}}\gamma } \right),
	\end{align}
	\begin{align}\label{cdf_snr}
		F\left( \gamma  \right) =1- \frac{1}{{\Gamma \left( m \right)}}\Gamma \left( {m,\frac{{m{\sigma ^2}}}{{\Omega {P_{{\rm{trans}}}}}}\gamma } \right),
	\end{align}
	where $ m $ denotes the shape parameter, $ \Omega $ is the spread-controlling parameter, $ \Gamma \left( x \right) $ is the gamma function \cite[Eq. (8.310.1)]{gradshteyn2007}, and $ \Gamma \left( {a,x} \right) $ is the upper  incomplete gamma function \cite[Eq. (8.350.2)]{gradshteyn2007}.

	\subsection{Overall Carbon Footprint Model}
AOLO is proposed as a framework for analyzing and optimizing low-carbon oriented wireless LLM services, where a comprehensive carbon footprint model is introduced to quantify the overall greenhouse gas emissions associated with serving one LLM inference request. 
The carbon footprint refers to the total amount of carbon dioxide and other greenhouse gas emissions, both direct and indirect, associated with the production, use, and disposal of a product or service throughout its entire life cycle \cite{wiedmann2008definition}.
The overall carbon footprint of the LLM inference service is determined by two primary phases: (i) the inference computation phase, with emissions from computational processes and the embodied carbon of the data center, and (ii) the wireless transmission phase, with emissions from data transmission and the embodied carbon of the base station. 
In the AOLO framework, the overall carbon footprint associated with serving one LLM inference request can be expressed as
\begin{align}
	{{\cal C}_{{\text{all}}}} = {{\cal C}_{\text{I}}} + {{\cal C}_{\text{C}}},
\end{align}
where ${{\cal C}_{\text{I}}}$ and ${{\cal C}_{\text{C}}}$ denote the carbon footprints of the inference computation phase and wireless communication phase, respectively. 

The above two phases can be analyzed in terms of their operational and embodied carbon footprints \cite{chien2023reducing,li2023carbon}. 
The operational carbon footprint refers to the direct emissions from the energy used during the active phases of the LLM inference computation  and wireless network transmission. 
The embodied carbon footprint refers to the emissions associated with the entire lifecycle of the hardware and infrastructure \cite{lovehagen2023assessing}, used in the data center for inference service and wireless network for transmission, from production and transportation to disposal.
			
	\section{Overall Carbon Footprint Optimization}\label{SEC_carbon_optimize}
    In this section, we detail the overall carbon footprint model in terms of inference computation and wireless communication phases, and formulate the carbon footprint optimization problem under constraints of QoE, energy consumption, and response times.
	
	\subsection{Inference Computation Carbon Footprint}
	
	For serving one LLM inference request, the carbon footprint of the inference computation phase ${{\cal C}_{\text{I}}}$ consists of the operational carbon footprint ${{\cal C}_{{\text{oper,I}}}}$, and the embodied carbon footprint ${{\cal C}_{{\text{emb,I}}}}$.

	The operational carbon footprint ${{\cal C}_{{\text{oper,I}}}}$ arises from the electric energy consumed by the data center during the serving of inference requests. It is calculated by multiplying the carbon intensity by the above energy consumption of the data center \cite{li2024toward,chien2023reducing}. Specifically, carbon intensity, measured in grams of CO2 emitted per unit of energy consumption, indicates the environmental impact or ``greenness" of the energy source. 
	Given that graphical processing units (GPUs) are the primary contributors to the energy consumption by computing hardware within the data center for deep learning applications \cite{buildcomputers2021,torbet2019}, we primarily focus on GPU electricity consumption to evaluate $ {{\cal C}_{{\text{oper,I}}}} $, aligning with common assumptions in recent studies \cite{chien2023reducing,dodge2022measuring}. 
	To incorporate additional sources of carbon footprint, we utilize power usage effectiveness (PUE) \cite{avelar2012pue}, a metric for assessing the energy efficiency of data centers. PUE is defined as the ratio of total energy usage in a data center, encompassing all auxiliary components such as cooling and power conversion, to the energy used by computing hardware \cite{henderson2020towards,lacoste2019quantifying}.
	Hence, we can model the operational carbon footprint of the  inference computation phase for one inference request as
	\begin{align}\label{OperaCarbonInfer}
		{{\cal C}_{{\text{oper,I}}}}={t_{{\text{infer}}}}{n_{\text{gpu}}}{P_{{\text{gpu}}}}{\eta }{\zeta _{1}},
	\end{align}
	where ${t_{{\text{infer}}}}$ is the inference  execution time of GPU,  $n_{\text{gpu}}$ is the number of GPUs used for this inference task, $P_{\text{gpu}}$ is the thermal design power of one GPU, ${\eta }$ is the PUE of data center\footnote{According to the Uptime Institute \cite{uptimeinstitute2023}, the global average PUE for data centers in 2023 is 1.58, whereas Google reports a fleet-wide PUE of 1.10 for its more efficient facilities in the same year \cite{google2024}.}, ${\zeta_{1}}$ is the carbon intensity of the data center, measured in gCO2/kWh.
 
	The embodied carbon footprint $ 	{{\cal C}_{{\text{emb,I}}}} $ is a portion of the data center’s total embodied carbon footprint, calculated based on the inference execution time $ t_{\text{infer}} $ relative to the overall operational lifespan of data center $ {T_{\text{DC}}} $ \cite{li2023toward}.	
	Here, we model the embodied carbon footprint of the inference computation phase for one inference request as
	\begin{align}\label{EmbCarbonInfer}
		{{\cal C}_{{\text{emb,I}}}} =\frac{t_{\text{infer}}}{T_{\text{DC}}} n_{\text{gpu}} {\cal C}_{{\text{GPU}}}^{{\text{E}}},
	\end{align}
    where $ {\cal C}_{{\text{GPU}}}^{{\text{E}}} $ is the embodied carbon footprint of the data center for per-GPU part (measured in kgCO2 per GPU), including the carbon footprint related to the manufacturing and packaging of computer hardware in the data center.

\begin{figure}[th]
	\centering
	\subfigure[\label{WordCount_InferTime}{Impact of inference output word count on the inference execution time with different LLMs.}]{\includegraphics[width=0.242\textwidth]{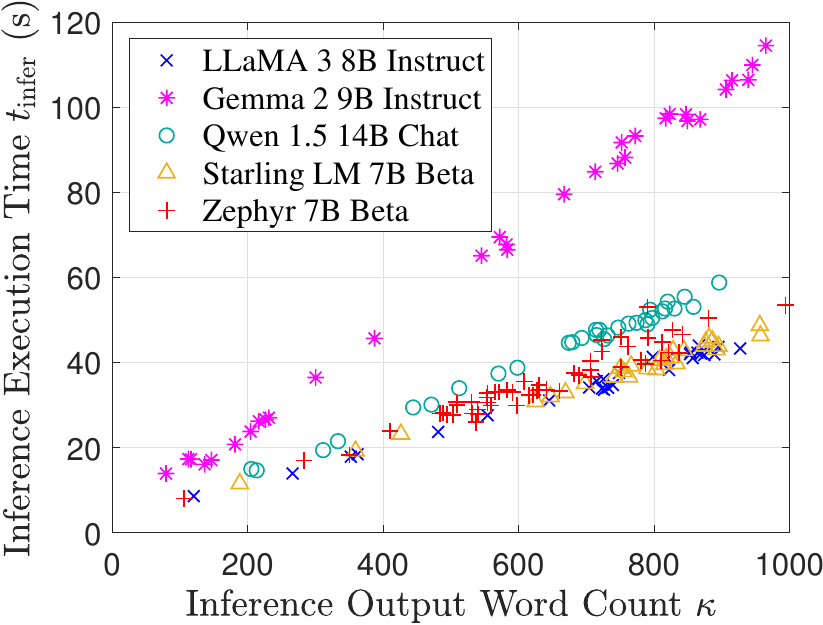}}
	\subfigure[\label{WordCount_Q}
	{The user's subjective QoE of the inference output with different word count.}]{\includegraphics[width=0.239\textwidth]{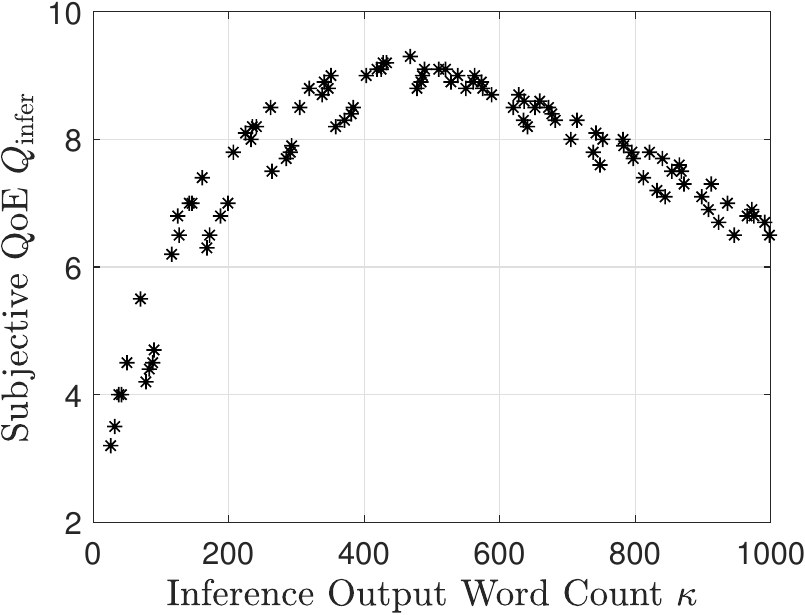}}
	\caption{Imapct of inference output word count on the inference execution time and subjective QoE. In subfigure (b), 100 stories with varying word counts (0-1000 words) on a shared theme are generated using \textit{Chat GPT-4o with canvas}. Then, \textit{Chat GPT-4} evaluates these stories from a five-year-old’s perspective, scoring them from 0 to 10 based on interestingness, clarity, and length suitability.}
	\label{fig-hebing-1}
\end{figure}

	The inference execution time  $t_{\text{infer}}$ is associated with the peak floating point operations per second (FLOP) capability of a single GPU $ {\Omega_{{\text{pF}}}} $ and the number of FLOP required by the inference request, and it can be expressed as
	\begin{align}\label{CarbonInferTime}
		{t_{{\text{infer}}}} = \frac{{{\psi _{{\text{OI}}}}{\psi _{{\text{IW}}}}}}{{{n_{\text{gpu}}}{\Omega _{{\text{pF}}}}}}{\kappa^\alpha },
	\end{align}%
	where $ {\psi _{{\text{OI}}}} $ is the FLOP count performed per inference operation, $ {\psi _{{\text{IW}}}} $ is the required inference operation count per output word, $\kappa$ is the word count of the inference output. 
	The parameter $ \alpha  \in \left( {0,1} \right] $ is a constant capturing the impact of various inference acceleration techniques used in LLMs, such as the key-value (KV) caching mechanism. These techniques effectively reduce computational overhead by leveraging intermediate results during inference, thereby affecting the scaling behavior captured by $ \alpha $.
	To further validate \eqref{CarbonInferTime}, the impact of inference output word count $ \kappa $ on the inference execution time $ t_{\text{infer}} $ across different LLMs\footnote{The LLMs presented are from KLU's  ``Best Open Source LLMs of 2024" \cite{klu2024opensource}.} is illustrated in Fig. \ref{fig-hebing-1} (a).


	Substituting \eqref{CarbonInferTime} into \eqref{OperaCarbonInfer} and \eqref{EmbCarbonInfer} and summing ${{\cal C}_{{\text{oper,I}}}}$ and ${{\cal C}_{{\text{emb,I}}}}$, the carbon footprint of the inference computation phase for one inference request is modeled as
	\begin{align}
{{{\cal C}_{\text{I}}} = \kappa^{\alpha} \omega {P_{{\text{gpu}}}}{\eta }{\zeta_1 }} + \kappa^{\alpha}  \omega \frac{{C_{{\text{GPU}}}^{{\text{E}}}}}{{{T_{{\text{DC}}}}}} ,
	\end{align}
	where  $\omega  = \frac{{{\psi _{{\text{OI}}}}{\psi _{{\text{IW}}}}}}{{{\Omega_{{\text{pF}}}} }}$ represents the time for one GPU to output 1 word of inference result (measured in GPU-sec per word).
		
	\subsection{Wireless Communication Carbon Footprint}
	For serving one inference request, the carbon footprint of the wireless communication phase ${{\cal C}_{\text{C}}}$ contains the operational carbon footprint ${{\cal C}_{{\text{oper,C}}}}$, and the embodied carbon footprint ${{\cal C}_{{\text{emb,C}}}}$.
	
	The operational carbon footprint ${{\cal C}_{{\text{oper,C}}}}$ originates from the electric energy used to transmit inference results and support the ongoing operation of the BS \cite{li2023carbon}. The latter consumption includes energy consumed for wireless network resource management and signal processing within the building baseband unit (BBU), as well as for cooling the hardware devices to ensure optimal operating temperatures at the BS. 
	Therefore, we can model the operational carbon footprint of the wireless communication phase for one inference request as
	\begin{align}\label{OperaCarbonCommun}
		{{\cal C}_{{\text{oper}},{\text{C}}}} = {t_{\text{trans}} } {P_{{\text{trans}}}}{\zeta_2 }  + \frac{{\kappa \beta {\zeta_2}}}{\mathcal{K}} {P_{{\text{fixed}}}},		
	\end{align}
	where $ t_{\text{trans}} $ is the time to transmit inference results, $ \beta $ is the average data size of each output word (measured in bits per word), $ {\zeta_2} $ is the local carbon intensity\footnote{The local carbon intensity can be obtained from sources such as Electricity Maps \cite{electricitymaps} and RiPiT \cite{ripit_uchicago}.} measured in gCO2/kWh, $ \mathcal{K} $ is the total data size that the BS transmits per second, $ P_{{\text{fixed}}} $ is the fixed power consumption required to maintain a BS, covering the power used by the BBU and the cooling equipment.

	The embodied carbon footprint ${{\cal C}_{{\text{emb}},{\text{C}}}}$ is calculated by apportioning the BS's total embodied carbon footprint according to the inference output data size relative to the total data size handled by the BS throughout its overall lifespan $T_{\text{BS}}$ \cite{ding2022carbon,lovehagen2023assessing}. We can model the embodied carbon footprint of the wireless communication phase for one inference request as 
	\begin{align}\label{EmbCarbonCommun}
		{{\cal C}_{{\text{emb}},{\text{C}}}}=\frac{{\kappa \beta }}{{\mathcal{K}{T_{{\text{BS}}}}}}{{\cal C}^{\text{E}}_{\text{BS}}},		
	\end{align}%
	where $ {{\cal C}^{\text{E}}_{\text{BS}}} $ is the total embodied carbon footprint of a single BS, which sums the carbon footprint of the production phase, transportation phase and construction phase \cite{ding2022carbon}.
	
	With the help of Shannon channel capacity, the wireless transmission time of inference results is expressed as 
	\begin{align}\label{CarbonCommunTime}
		t_{\text{trans}} =  \frac{\kappa \beta}{B{\log _2}\left( {1 + \gamma } \right)},
	\end{align}
    where $ B $ is the channel bandwidth.
     Substituting \eqref{CarbonCommunTime} into \eqref{OperaCarbonCommun} and summing ${{\cal C}_{{\text{oper,C}}}}$ and ${{\cal C}_{{\text{emb,C}}}}$, the carbon footprint of the wireless communication phase for one inference request is modeled as
	\begin{align}\label{eq_Carbon_trans_1}
	{{\cal C}_{\text{C}}} = \frac{{\kappa \beta \zeta_2 {P_{{\text{trans}}}}}}{{B{{\log }_2}\left( {1 + \gamma } \right)}} + \frac{{\kappa \beta \left( {\zeta_2 {P_{{\text{fixed}}}}{T_{{\text{BS}}}} + {{\cal C}^{\text {E}}_{\text{BS}}}} \right)}}{{\mathcal{K}{T_{{\text{BS}}}}}}.
	\end{align}

The outage event, when the instantaneous SNR approaches zero with non-zero probability, is considered for the wireless transmission of inference results. 
To prevent excessive transmission delay and inefficient energy usage, the transmission is terminated when the instantaneous SNR falls below a predefined threshold $ \gamma_{\text{th}} $.
With the help of \eqref{cdf_snr}, the outage probability of wireless transmission in our considered system can be expressed as
\begin{align}\label{eq:op}
	P_{\text{op}}=Pr \left[ {\gamma  < {\gamma _{{\text{th}}}}} \right] = F\left( {{\gamma _{{\text{th}}}}} \right) =1- \frac{\Gamma \left( {m,\frac{{m{\sigma ^2}}}{{\Omega {P_{{\text{trans}}}}}}{\gamma _{{\text{th}}}}} \right)}{{\Gamma \left( m \right)}}.
\end{align}
The reliability of communication is ensured when the outage probability satisfies $ P_{\text{op}} < \epsilon $, where the threshold $ \epsilon $ is arbitrarily close to 0. 
The threshold $ \epsilon $ serves as a critical system design parameter for achieving both reliable and sustainable wireless communication.
As we can observe in \eqref{eq:op}, there exists a one-to-one correspondence between $ {\gamma _{{\text{th}}}} $ and $ \epsilon $, which is given by
\begin{align}\label{eq_realationship}
	1-\epsilon = \frac{\Gamma \left( {m,\frac{{m{\sigma ^2}}}{{\Omega {P_{{\text{trans}}}}}}{\gamma _{{\text{th}}}}} \right)}{{\Gamma \left( m \right)}}.
\end{align}

The average transmission time and the average carbon footprint of the wireless communication phase are determined as
\begin{align}
	\label{eq_t_trans_14}
	{{\bar t}_{{\text{trans}}}} &=\frac{1}{{1 -\epsilon }} \int_{\gamma_{\text {th}}}^\infty  {{t_{{\text{trans}}}}f\left( \gamma  \right){\rm d}\gamma},\\
    \label{eq_Cc_trans_15}
	{{\bar {\cal C}}_{\text{C}}} &=\frac{1}{{1 -\epsilon }} \int_{\gamma_{\text {th}}}^\infty  {{{\cal C}_{\text{C}}}f\left( \gamma  \right)} {\rm d}\gamma , 
\end{align}
where the value of $ \gamma_{\text{th}} $  is determined by the predefined system design parameter $ \epsilon $ using \eqref{eq_realationship}.
With the help of \eqref{pdf_snr}, \eqref{CarbonCommunTime} and \eqref{eq_Carbon_trans_1}, we can obtain the expressions of average transmission time $ {{\bar t}_{{\text{trans}}}} $ and average carbon footprint $ {{\bar {\cal C}}_{\text{C}}} $ of the wireless communication phase, which is presented in the following theorem.
\begin{theorem}\label{The_averageT}
The closed-form average transmission time and average carbon footprint of the wireless communication phase in our considered system are derived as
\begin{align}\label{eq_t_trans_16}
&{{\bar t}_{{\rm{trans}}}}  =\frac{1}{{1 -\epsilon }} {\left( {\frac{{m{\sigma ^2}}}{{\Omega P_{{\rm{trans}}}}}} \right)^{\!\!m}}\!\frac{{\kappa \beta }}{{\Gamma \left( m \right)B}}{I_c},\\
\label{eq_Cc_17}
&{{\bar {\cal C}}_{\rm{C}}}\! =\!\frac{1}{{1 \!-\!\epsilon }} \frac{{\kappa \beta \zeta_2 {m^m}{\sigma^{2m}}}}{{\Gamma \!\left( m \right){\Omega ^m}P_{{\rm{trans}}}^{m - 1}B}}{I_c}\! +\! \frac{{\kappa \beta \!\left( {\zeta_2 {P_{{\rm{fixed}}}}{T_{{\rm{BS}}}} \!+\! {{\cal C}^{\rm{E}}_{\rm{BS}}}} \right)}}{{(1 -\epsilon )\mathcal{K}{T_{{\rm{BS}}}}}}, 
\end{align}
where 
\begin{align}\label{eq_Ic}
{I_c} &\!=\!  - 4{\pi ^2}\ln\! \left( 2 \right)\gamma _{{\rm{th}}}^m \notag \\
&\times\! G_{1,1:2,2:1,0}^{0,1:0,1:0,1}\!\left(\!\!\! {\left. {\begin{array}{*{20}{c}}
			{ - m}\\
			{ - m\! +\! 1}
	\end{array}} \!\!\right|\!\!\left. {\begin{array}{*{20}{c}}
			{0,1}\\
			{1,1}
	\end{array}} \!\right|\!\left. {\begin{array}{*{20}{c}}
			1\\
			- 
	\end{array}}\! \right|\frac{1}{{{\gamma _{{\rm{th}}}}}},\frac{{\Omega {P_{{\rm{trans}}}}}}{{m{\sigma ^2}{\gamma _{{\rm{th}}}}}}} \right),
\end{align}
and $G_{{p_1},{q_1}:{p_2},{q_2}:{p_3},{q_3}}^{{m_1},{n_1}:{m_2},{n_2}:{m_3},n_3}\left(  \cdot  \right) $ is the extended generalized bivariate Meijer G-function (EGBMGF) \cite[07.34.21.0081.01]{webWolfram}.
\end{theorem}
\begin{IEEEproof}
	Please refer to Appendix \ref{Appendix_average}.
\end{IEEEproof}

Note that the EGBMGF in \eqref{eq_Ic} can be efficiently implemented in Matlab \cite{chergui2016performance} and Mathematica \cite{ansari2011new}. 

\subsection{Problem Formulation}\label{Problem_Formulation}

To achieve a low carbon footprint in the considered wireless network-aided LLM service system, an optimization problem is formulated in this part. The objective is to minimize the overall carbon footprint associated with serving one LLM inference request, by optimally adjusting the transmit power $ P_{\text{trans}} $, and the inference output word count $ \kappa $. 
Let $ Q_{\text{infer}} $ denote the user's subjective QoE of the inference result, which serves as a measure of inference quality. 
To ensure user's subjective QoE, $ Q_{\text{infer}} $ must meet or exceed a predefined threshold  $ Q_{\text{infer}}^{\text{th}} $. 
The energy consumption for each inference request, determined by $ P_{\text{trans}} $ and $ \kappa $, should not exceed the upper limit $ E_{\text{th}} $. 
Additionally, the inference execution time $ t_{\text{infer}} $ and the wireless transmission time $ t_{\text{trans}} $  should be within the upper limits $ t_{\text{infer}}^{{\text{th}}} $ and $ t_{\text{trans}}^{{\text{th}}} $, respectively.
The transmit power should be limited within an upper boundary $ P_{{\text{trans}}}^{\max } $. 
Thus, the optimization problem to minimize the overall carbon footprint can be formulated as 
\begin{subequations}
	\label{problemCarbon}
	\begin{align}
		\mathop {\min }\limits_{\left\{ {\kappa ,{P_{{\text{trans}}}}} \right\}}& \quad {{\cal C}_{{\text{I}}}}+{{\bar {\cal C}}_{\text{C}}} 	\tag{\ref{problemCarbon}}	
		\\		
		{\rm s.t.}& \quad Q_{\text{infer}} \ge {Q_{\text{infer}}^{{\text{th}}}}, \tag{\ref{problemCarbon}a}\\
		& \quad {\rho_1}\kappa  + {\rho_2}{P_{{\text{trans}}}} \le {E_{{\text{th}}}},  \tag{\ref{problemCarbon}b}\\
		& \quad {t_{\text{infer}}} \le t_{\text{infer}}^{{\text{th}}},	\tag{\ref{problemCarbon}c}\\
		& \quad {{\bar t}_{{\text{trans}}}} \le t_{\text{trans}}^{{\text{th}}}, \tag{\ref{problemCarbon}d}	\\
		& \quad {P_{{\text{trans}}}} \le P_{{\text{trans}}}^{\max }, \tag{\ref{problemCarbon}e}
	\end{align}
\end{subequations}
where $ {\rho_1} $ and $ {\rho_2} $ are coefficients of $ P_{\text{trans}} $ and $ \kappa $, characterizing the energy consumption of serving an inference request.

The user’s subjective QoE is influenced by various factors, such as the GenAI model’s capabilities to understand user intent, and the presence of hallucinations in responses. Without loss of generality, we focus on the impact of output text length on QoE, as overly short texts may fail to adequately develop a narrative, while excessively long texts can cause users to lose focus. To validate this, we designed an experiment where 100 stories with varying word counts (0–1000 words) were generated using \textit{ChatGPT-4o with canvas}. To quantify QoE, \textit{ChatGPT-4} model simulated a five-year-old kid's perspective, evaluating each story based on interestingness, ease of understanding, and length suitability, assigning a $Q_{\text{infer}}$ score (0–10). As shown in Fig. \ref{fig-hebing-1} (b), $Q_{\text{infer}}$ initially increases with word count $ \kappa $ due to richer narratives but declines when $ \kappa $ becomes too large, as overly long stories hinder attention retention. This demonstrates the importance of optimizing output length to balance engagement and focus.

To address the optimization problem formulated in \eqref{problemCarbon}, the SDRL algorithm is proposed in Section \ref{SEC_SDRL}, which is designed to minimize the overall carbon footprint while meeting the constraints of QoE, energy consumption, and response times.

\section{Spiking Neural Network-based Deep Reinforcement Learning}\label{SEC_SDRL}

Before delving into the SDRL algorithm, we present the motivations and preliminaries of adopting SNN as the actor network. 
Then the SDRL algorithm is proposed to optimize the overall carbon footprint of the LLM inference service by adjusting the inference output word count $ \kappa $ and the transmit power $ P_{{\text{trans}}} $.

\begin{figure*}[t]
	\centering
	\includegraphics[width=0.8\textwidth]{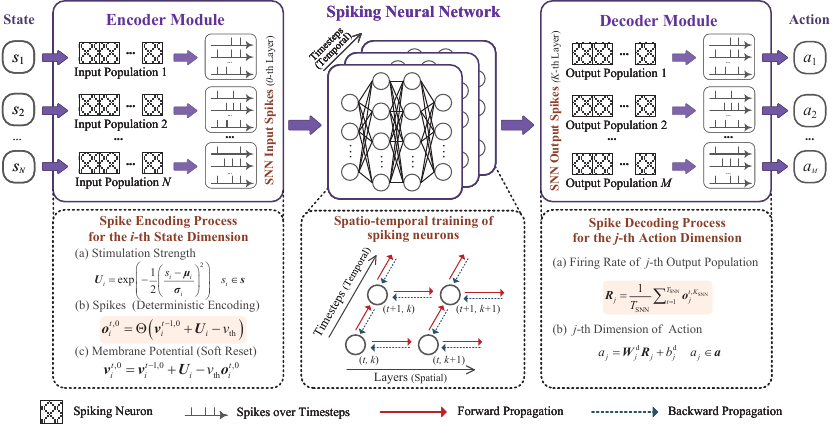}
	\caption{Illustration of SNN-based actor network employing PopSAN method, tailored to generate optimal decisions for minimizing the overall carbon footprint associated with serving an inference request. 
	The encoder transforms each state dimension into spiking activity employing population coding, generating input spikes for the SNN. The SNN module processes these spikes through its spatio-temporal structure to produce output spikes. Finally, the decoder calculates the firing rates of output populations and converts them into continuous action values, completing the decision-making process. }
	\label{FIG:PopSAN}
\end{figure*}

\subsection{Motivations and Preliminaries of Adopting SNN}

DRL has gained popularity for solving complex, high-dimensional tasks, but the dense computations of MLP-based networks lead to high energy consumption, which limits their scalability in energy-constrained settings. In contrast, SNNs, with their event-driven computation inspired by biological systems, offer a promising, energy-efficient alternative.
The exceptional energy efficiency of SNNs aligns naturally with the goal of minimizing the carbon footprint of LLM inference services. To leverage this advantage, the actor network in the proposed SDRL algorithm is implemented using the population-coded spiking actor network (PopSAN) \cite{tang2021deep}, which is characterized by population coding.
By encoding each dimension of the state and action spaces into the activities of individual input and output spiking neuron populations, the representation capacity of spiking neurons is significantly enhanced. 
This improvement enables PopSAN to achieve performance levels comparable to, or even exceeding, those of traditional MLP-based actor networks, while significantly reducing energy consumption. 
As a result, it serves as an ideal choice for addressing complex optimization problems with low-carbon and energy-efficiency demands.

With the help of the PopSAN method, the SNN-based actor network comprises three main components: a population encoder, an SNN module, and a population decoder.
The population encoder transforms each dimension of the state into the spiking activity of an individual neuron population. During forward propagation, these input populations drive the SNN module to generate output spikes, which are then decoded into the corresponding actions after every $T_{\text{SNN}}$ spiking timesteps.
The architecture of the SNN-based actor network designed to generate optimal decisions for minimizing the overall carbon footprint associated with serving one LLM inference request is illustrated in Fig. \ref{FIG:PopSAN}.

\subsubsection{Population Encoder Module}
Considering the limited representation capacity of a single spiking neuron, the population encoding is performed to translate the $ N $-dimensional state $ \boldsymbol{s} $ into spikes by $ N $ individual input populations \cite{tang2021deep}. 
In each population, the encoder population dimension is defined by the count of spiking neurons $ N_{\text{e}} $.
The encoded state information is fed into the SNN module. 
For the $ i $-th state dimension, the population encoding process, where the encoder module encodes the state $ s_i \in \boldsymbol{s}, i \in \left\{ {1,...,N} \right\}  $, into spikes by the $ i $-th input population, can be divided into two phases.
Firstly, the value of state $ s_i  $ is transformed to the simulation strength $ {\boldsymbol{U}_i}  $ of each neuron in the $ i $-th input population with Gaussian receptive fields $ \left( {\boldsymbol{\mu}_i ,\boldsymbol{\sigma}_i } \right) $, which can be formulated as
\begin{align}\label{eq:simulation_strength}
{\boldsymbol{U}_i} = \exp \left( { - \frac{1}{2}{{\left( {\frac{{{s_i} - \boldsymbol{\mu}_i }}{\boldsymbol{\sigma}_i }} \right)}^2}} \right), 
\end{align}
where $ \boldsymbol{\mu}_i $ is initialized 
with a uniform distribution over the state space, $ \boldsymbol{\sigma}_i $ is preset to be large enough to guarantee non-zero population activity across the entire state space \cite{tang2021deep}, and both of them are task-specific trainable parameters.
Secondly, the simulation strength $ {\boldsymbol{U}_i} $ is utilized to generate the neuron spikes of the $ i $-th input population. By employing deterministic encoding, the neurons are modeled as the one-step soft-reset integrate-and-fire (IF) structure \cite{burkitt1999analysis}, where the neurons receive presynaptic inputs denoted by  $ {\boldsymbol{U}_i} $. 
The dynamics of the neurons in the $ i $-th input population can be described as
\begin{align}\label{eq:o_1}
	\boldsymbol{o}_i^{t,0} &= \Theta \left( {\boldsymbol{v}_i^{t - 1,0} + {\boldsymbol{U}_i} - {v_{{\text{th}}}}} \right),\\
	\label{eq:v_1}
	\boldsymbol{v}_i^{t,0} &= \boldsymbol{v}_i^{t - 1,0} + {\boldsymbol{U}_i} - {v_{{\text{th}}}}\boldsymbol{o}_i^{t,0},
\end{align}
where $ \boldsymbol{o}_i^{t,0} $ and $ \boldsymbol{v}_i^{t,0} $ are the spikes and the membrane potential of the neurons from the $ i $-th input population at spiking timestep $ t \in \left\{ {1,...,T_{\text{SNN}}} \right\}  $, $ {v_{{\text{th}}}} $ is the threshold potential, and $ \Theta \left( \cdot \right) $ is the Heaviside step function \cite{bracewell1966fourier}, which is given by $ \Theta(x) = 1 $ for $ x \ge 0 $ and $ \Theta(x) = 0 $ for $ x < 0 $.

\subsubsection{SNN Module}
The current-based leaky integrate-and-fire (LIF) model is adopted for the spiking neurons to construct the fully connected SNN module \cite{memmesheimer2014learning,gerstner2014neuronal}, where the spike information is transmitted across $ K_{\text{SNN}} $ layers and iterates over $ T_{\text{SNN}} $ spiking timesteps. 
There are three phases for the dynamics of LIF-based spiking neurons in the $ k $-th layer, $ k \in \left\{ {1,...,K_{\text{SNN}}}  \right\}  $, at spiking  timestep $ t $. First, the presynaptic spikes $ {\boldsymbol{o}^{t - 1,k}} $ is integrated into the current. Then, the current $ {\boldsymbol{c}^{t,k}} $ is integrated into the membrane potential. 
Subsequently, the spiking neuron fires a spike $ {\boldsymbol{o}^{t,k}} $ when its membrane potential $ {\boldsymbol{v}^{t,k}} $ exceed the predetermined threshold $ v_{th} $. The overall spike information processing can be mathematically formulated as 
\begin{align}
	\label{eq-LIFupdate_c}
	{\boldsymbol{c}^{t,k}} & = {d_c}{\boldsymbol{c}^{t - 1,k}} + {\boldsymbol{W}^k}{\boldsymbol{o}^{t,k - 1}} + {\boldsymbol{b}^k},\\
	\label{eq-LIFupdate_v}
	{\boldsymbol{v}^{t,k}} & = {d_v}{\boldsymbol{v}^{t - 1,k}}\left( {1 - {\boldsymbol{o}^{t - 1,k}}} \right) + {\boldsymbol{c}^{t,k}},\\
	\label{eq:o_2}
	{\boldsymbol{o}^{t,k}} & = \Theta \left( {{\boldsymbol{v}^{t,k}} - {v_{\text{th}}}} \right),
\end{align}
where $ d_c $ and $ d_v $ are the current and membrane potential decay factors, and $ {\boldsymbol{W}^k} $ and $ {\boldsymbol{b}^k} $ is the weight and bias of the $ k $-th layer of SNN.

\subsubsection{Population Decoder Module}
The spiking neuron activity generated at the SNN output layer is converted into the $ M $-dimensional action $ \boldsymbol{a} $ by the population decoder module, where the output layer, i.e., the $ K_{\text{SNN}} $-th layer, comprises of $ M $ output populations. 
In each population, the decoder population dimension is defined by the count of spiking neurons $ M_{\text{d}}$. 
For the $ j $-th action dimension, there are two phases for the decoder module to decode the activity of $ j $-th output population into the corresponding action $ a_j \in \boldsymbol{a}, j \in \left\{ {1,...,M}  \right\} $. 
First, the spikes generated from the $ j $-th output population are summed up over $ T_{\text{SNN}} $ to obtain the firing rate $ \boldsymbol{R}_j $. Second, the corresponding action $ a_j $ is calculated by the weighted sum of the firing rate $ \boldsymbol{R}_j $. The decoding process of spikes from the $ j $-th output population can be formulated as 
\begin{align}
\label{eq:firingRate}
{\boldsymbol{R}_j} &= \frac{1}{T_{\text{SNN}}}\sum\nolimits_{t = 1}^{T_{\text{SNN}}} {\boldsymbol{o}_j^{t,K_{\text{SNN}}}},\\
\label{eq:a_j}
{a_j} &= \boldsymbol{W}_j^{\text{d}} {\boldsymbol{R}_j} + b_j^{\text{d}},
\end{align}
where $ \boldsymbol{W}_j^{\text{d}} $ and $ b_j^{\text{d}} $ are the weight and bias of the decoder module for the $ j $-th action dimension.

\subsubsection{Training}
Considering the discontinuity of spikes utilized in SNN, the regular backpropagation of DNN cannot be directly applied to train multilayered SNNs, which have the advantage of handling spatiotemporal information. 
The SNN parameters are updated with the help of the extended spatio-temporal backpropagation as introduced in \cite{tang2020reinforcement}.
Since the Heaviside step function that defines a spike in \eqref{eq:o_1} and \eqref{eq:o_2} is non-differentiable, a pseudo-gradient function is required to approximate the spike gradient. 
The rectangular function is adopted as the pseudo-gradient function, which demonstrates the best empirical performance in \cite{wu2018spatio}.
The parameters in SNN are updated every $ T_{\text{SNN}} $ spiking timesteps.

\subsection{MDP Formulation}
The decision-making process in the carbon footprint optimization problem is modeled as a Markov decision process (MDP), comprising a state space, an action space, and a reward function. Each element is described in detail below.

\subsubsection{State Space}
The state space $ {\cal S} $ is the set of all possible states that define the environment's configurations, representing the information available to the agent at any given time to make decisions. 
The state $ \boldsymbol{s} \in {\cal{S}} $ consists of five components, i.e., the shape parameter $ m $, the spread-controlling parameter $ \Omega $, the channel bandwidth $ B $, the carbon intensity of the data center $ {\zeta _1} $, and the local carbon intensity $ {\zeta _2} $. Accordingly, $ \boldsymbol{s} $ is defined as 
\begin{align}
{\boldsymbol{s}} = \left\{ {m,\Omega ,B,{\zeta _1},{\zeta _2}} \right\} \in {\cal S}.
\end{align}

\subsubsection{Action Space}
The action space $ \boldsymbol{A} $ contains all possible actions that the agent can take in the given environment. The action $ {\boldsymbol{a}} \in {\cal A} $ is composed of two components: the inference output word count $ \kappa $, which affects the complexity and quality of the output, and the transmit power $ P_{{\text{trans}}} $, which impacts energy consumption during transmission. 
Together, these two components affect the balance between the quality of LLM inference services, energy consumption, and carbon efficiency.
The action $ {\boldsymbol{a}} $ is defined as  
\begin{align}
{\boldsymbol{a}} = \left\{ {\kappa ,{P_{{\text{trans}}}}} \right\} \in {\cal A}.
\end{align}

\subsubsection{Reward Function}
The reward function maps a state $ \boldsymbol{s} $ and an action $ \boldsymbol{a} $ to a scalar value $ r $, providing feedback to the agent to guide its learning and decision-making process.
In our optimization problem, a reward function is designed based on the optimization objective presented in \eqref{problemCarbon} to minimize the overall carbon footprint for the LLM inference service. 
In addition, a penalty mechanism is incorporated to ensure all predefined constraints are satisfied, including the inference quality, energy consumption, and time limits for inference and transmission. If any constraint is violated, the reward is assigned a penalty of $ -100 $, discouraging infeasible actions. 
Consequently, the policy is trained through interactions with the environment by maximizing the reward function, which is defined as
\begin{align}\label{eq:reward}
	r &= 
	\begin{cases}
		-( C_{\text{I}} + \overline{C}_{\text{C}}), & \text{if all constraints are satisfied}, \\
		-100, & \text{if any constraint is violated}.
	\end{cases}
\end{align}

\subsection{Algorithm Architecture}
\begin{figure}[t]
	\centering
	\includegraphics[width=0.44\textwidth]{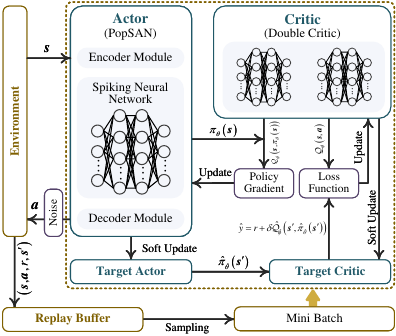}
	\caption{Overall architecture of the SDRL algorithm with an SNN-based actor network utilizing PopSAN method and a double-critic network. }
	\label{FIG:DDPG}
\end{figure}
The overall architecture of SDRL, depicted in Fig. \ref{FIG:DDPG}, includes six components: an SNN-based actor network, a double-critic network, a target actor, a target critic, the replay buffer, and the environment, which together optimize the policy.

\begin{algorithm}[ht]
	{\label{algorithm1}
		\caption{SDRL Algorithm for Overall Carbon Footprint Optimization of LLM Inference Service}
		Initialize the network parameters $ \theta $, $ \phi $, the target network parameters  $ \hat \theta  \leftarrow \theta  $, $ \hat \phi  \leftarrow \phi $, the replay buffer $ \varepsilon $, and configure all hyperparameters\;
		\For{episode $ =1 $ to $ H $}
		{\For{timestep $l =1 $ to $ L $}{
				Observe the environment to obtain the state  $ \boldsymbol{s} = \left\{ {m,\Omega ,B,{\zeta _1},{\zeta _2}} \right\} $\;
				\For{$ t=1 $ to $ T_{\text{SNN}} $}{
					\For{$ i=1 $ to $ N $}{
						Calculate the simulation strength $ \boldsymbol{U}_i $ of the $ i $-th component in $ \boldsymbol{s}$ by \eqref{eq:simulation_strength}\;
						Obtain the spikes $ \boldsymbol{o}_i^{t,0} $ and update the membrane potential $ \boldsymbol{v}_i^{t,0} $ of the $ i $-th input population by \eqref{eq:o_1} and \eqref{eq:v_1}\;}
					\For{$ k=1 $ to $ K_{\text{SNN}} $}{
						Update the spiking neurons in the $ k $-th layer at spiking timestep $ t $ by \eqref{eq-LIFupdate_c}, \eqref{eq-LIFupdate_v} and \eqref{eq:o_2}\;}
				}
				\For{$ j=1 $ to $ M $}{
					Calculate the firing rate $ \boldsymbol{R}_j $ of the $ j $-th output population by \eqref{eq:firingRate}, and obtain the $ j $-th component in $ \boldsymbol{a} $ by \eqref{eq:a_j}\;}
				Obtain the optimal decisons $ {\boldsymbol{a}} = \left\{ {\kappa ,{P_{{\text{trans}}}}} \right\} $\;
				Receive the reward $ r $ according to \eqref{eq:reward} and transition to the next state $ \boldsymbol{s}' $\;
				Store $ \left( {\boldsymbol{s},\boldsymbol{a},r,\boldsymbol{s}'} \right) $ into the replay buffer $ \varepsilon $\;
				Randomly sample a batch of transitions $ \mathcal{B}={\left( {\boldsymbol{s},\boldsymbol{a},r,\boldsymbol{s}'} \right)} $ from the replay buffer $ \varepsilon $\;
				Update the policy parameter $ \theta $ and Q-function parameter $ \phi $ using $ \mathcal{B} $ by \eqref{eq:theta_update} and \eqref{eq:phi_update}\;
				Update the target network parameters $ \hat \theta $ and $ \hat \phi $ by \eqref{eq-targetupdate-theta} and \eqref{eq-targetupdate-phi}, respectively\;
		}}
	}
\end{algorithm}

\subsubsection{SNN-based Actor Network}
Using the PopSAN method \cite{tang2021deep}, the actor network $ {\pi _\theta } $, parameterized by $ \theta $, is constructed based on an SNN. 
The input to the actor network is the current state $ \boldsymbol{s} $, and its output is a deterministic action $ \boldsymbol{a} = {\pi _\theta }\left( \boldsymbol{s} \right) $, which is optimized by maximizing the Q-value estimated by the double-critic network. 
The dimensions of the state $ \boldsymbol{s} $ and the action $ \boldsymbol{a} $ determine the number of individual input populations $ N $ for the encoder module and individual output populations $ M $ for the decoder module, respectively.
Policy improvement is achieved by updating the actor network. The policy optimization objective is maximizing the expectation of the Q-value estimated from $ {{\cal Q}_\phi } $ over the mini-batch of transitions of size $ |\mathcal{B}| $, which is defined as,
\begin{align}
	{\cal J}\left( \theta  \right) = {\mathbb{E}_{\boldsymbol{s}\sim{\cal B}}}\left[ {{{\cal Q}_\phi }\left( {\boldsymbol{s},{\pi _\theta }\left( \boldsymbol{s} \right)} \right)} \right].
\end{align}
The parameter $ \theta $ of the SNN-based actor network is updated using policy gradient ascent, which is formulated as,
\begin{align}
	\label{eq:theta_update}
	\theta  \leftarrow \theta  + {\lambda _a}{\mathbb{E}_{\boldsymbol{s}\sim{\cal B}}}\left[ {{\nabla _{\boldsymbol{a}}}{{\cal Q}_\phi }{{\left. \left( {\boldsymbol{s},\boldsymbol{a}} \right)\right|}_{\boldsymbol{a} = {\pi _\theta }\left( \boldsymbol{s} \right)}}     {\nabla _\theta }{\pi _\theta }\left( \boldsymbol{s} \right)} \right],
\end{align}
where $ \lambda _a $ is the learning rate of the actor network.
To encourage exploration in the continuous action space, Gaussian noise $ \mathcal{N}(0,\sigma_0^2) $ is added to the actor network's output during interaction with the environment. 
Moreover, to enhance the learning stability, a target actor network $ {{\hat \pi }_{\hat \theta }} $, parameterized by $ {\hat \theta } $, is deployed, sharing the same structure as $ {\pi _\theta } $.

\subsubsection{Double-Critic Network and Target Networks}
An MLP-based double-critic network $ {{\cal Q}_\phi } $, parameterized by $ \phi  = \left\{ {{\phi _1},{\phi _2}} \right\} $, is employed in SDRL to estimate the Q-values of state-action pairs, where two separate critic networks, $ {{\cal Q}_{{\phi _1}}} $ and $ {{\cal Q}_{{\phi _2}}} $, are updated independently. 
During training, the double-critic network takes $ \boldsymbol{s} $ and $ \boldsymbol{a} $ as inputs, and outputs the smaller one of the two Q-value estimates from $ {{\cal Q}_{{\phi _1}}} $ and $ {{\cal Q}_{{\phi _2}}} $, i.e., $ {{\cal Q}_\phi }\left( {\boldsymbol{s},\boldsymbol{a}} \right) = \min \left\{ {{{\cal Q}_{{\phi _1}}}\left( {\boldsymbol{s},\boldsymbol{a}} \right),{{\cal Q}_{{\phi _2}}}\left( {\boldsymbol{s},\boldsymbol{a}} \right)} \right\}$.
This approach reduces overestimation bias, and stabilizes policy learning by providing a conservative Q-value estimation.
Accurate Q-value estimates for state-action pairs are  essential for guiding the policy in optimizing decisions to maximize long-term rewards. 
To improve the accuracy of Q-value estimation, we update the Q-function by minimizing the expectation of temporal difference (TD) error over the  mini-batch of transitions of size $ |\mathcal{B}| $. Thus, the loss function is expressed as,
\begin{align}
	{\cal L}\left( {{\phi _1},{\phi _2}} \right)={\mathbb{E}_{\left( {\boldsymbol{s},\boldsymbol{a},r,\boldsymbol{s}'} \right)\sim {\cal B}}}\left[ {\sum\limits_{r = 1,2} {\left( {\hat y - {{\cal Q}_{{\phi _r}}}\left( {\boldsymbol{s},\boldsymbol{a}} \right)} \right)^2} } \right],
\end{align}
where $ \hat y = r + \delta {{\hat {\cal Q}}_{\hat \phi }}\left( {\boldsymbol{s}',{{\hat \pi }_{\hat \theta }}\left( {\boldsymbol{s}'} \right)} \right)$ is the target Q value, and $ \delta $ is the discount factor for future rewards. The parameters $ \phi_1 $ and $ \phi_2 $ in the double critic networks are updated with the gradient descent method as,
\begin{align}
	\label{eq:phi_update}
	{\phi _r} \leftarrow {\phi _r} - {\lambda _c}{\nabla _{{\phi _r}}}{\cal L}\left( {{\phi _1},{\phi _2}} \right),\quad r \in \left\{ {1,2} \right\},
\end{align}
where $ \lambda _c $ is the learning rate of the double-critic network.
The improved Q-function provides more reliable feedback to support the actor network in learning an optimal policy.
Similarly, a target double critic network $ {{\hat {\cal Q}}_{\hat \phi }} $, parameterized by $ {\hat \phi } $, is adopted.

The target networks, $ {{\hat \pi }_{\hat \theta }} $ and $ {{\hat {\cal Q}}_{\hat \phi }} $, are employed to ensure stability during training, by stabilizing both policy updates and Q-value estimates. 
While the online networks undergo frequent gradient ascent and descent, the  target networks remain static. 
The corresponding parameters $ {\hat \theta } $ and $ {\hat \phi } $ are gradually updated through a soft update mechanism, which is given by
\begin{align}
	\label{eq-targetupdate-theta}
	\hat \theta  \leftarrow \tau \theta  + \left( {1 - \tau } \right)\hat \theta,\\
	\label{eq-targetupdate-phi}
	\hat \phi  \leftarrow \tau \phi  + \left( {1 - \tau } \right)\hat \phi,
\end{align}
where $ \tau \in (0,1] $ is the soft update rate for target networks. 
A smaller $ \tau $ slows down the updates, which improves training stability but lengthens the overall training process. By adjusting $ \tau $, the target network stability and learning speed can be balanced.

\subsubsection{Replay Buffer}
A replay buffer $ \varepsilon $ is adopted to break up the temporal correlations of the training samples through random sampling. 
During interaction with the environment, the transition tuple $ \left( {\boldsymbol{s},\boldsymbol{a},r,\boldsymbol{s}'} \right) $ is stored in the replay buffer, where $ \boldsymbol{s}' $ is the next-timestep state generated by taking action $ \boldsymbol{a} $ in the real environment. 
After the preliminary phase of environment exploration, a mini-batch of transitions, denoted by $ \mathcal{B} $, is randomly sampled from the replay buffer to update the actor and critic networks together.

The pseudocode of our proposed SDRL algorithm for the overall carbon footprint optimization of LLM inference service is provided in Algorithm \ref{algorithm1}.

\section{Performance Evaluation}\label{sec:evaluation}
In this section, we outline the simulation setup, and evaluate the performance of our proposed SDRL algorithm.

\subsection{Simulation Setup}
\subsubsection{Simulation Parameters}
The LLM inference service is hosted in a data center, where the carbon intensity is in the range of $ \zeta_1 =[50,150] $ gCO2/kWh and the PUE is $ \eta=1.58 $.
The wireless transmission of the inference output is handled by a base station, where the local carbon intensity is $ \zeta_2=[400,900] $ gCO2/kWh and the outage probability threshold is $ \epsilon =0.1 $. 
The other simulation parameter settings for the wireless transmission and the data center supporting the LLM inference service are summarized in Table \ref{tablexxx}.
The simulations are conducted on a platform with an NVIDIA GeForce RTX 4090 GPU, utilizing PyTorch 2.4.1 and Python 3.8.2.

\begin{table}[t]
	\centering
	\caption{Parameters used in simulations \cite{chien2023reducing,ding2022carbon}}
	{\small\begin{tabular}{|c|c|}
		\hline
		\rowcolor[rgb]{ .949,  .949,  .949} \textbf{Parameters} & \textbf{Value} \bigstrut\\
		\hline
		\hline
		$B$, $m$, $\Omega$ & [15, 25] MHz, [2, 12], [0.1, 10] \bigstrut[t]\\
		\rowcolor[rgb]{ .949,  .949,  .949} $ P_{{\text{fixed}}} $, $ \mathcal{K} $ & 600 W, 1000 Mbps \\
		$\alpha$, $\beta$, $P_{\text gpu}$, $n_{\text gpu}$ & 0.8, 50 bits, 0.428 kW, 8 \\
		\rowcolor[rgb]{ .949,  .949,  .949} $ {\psi _{{\text{OI}}}} $, $ {\psi _{{\text{IW}}}} $, $ {\Omega_{{\text{pF}}}} $ & 0.35 TFLOPs, 5, 156 TFLOPs \\
		$ {\cal C}_{{\text{GPU}}}^{{\text{E}}} $, $ {{\cal C}^{\text {E}}_{\text{BS}}} $ & 318 kgCO2, 6500 kgCO2 \\
		\rowcolor[rgb]{ .949,  .949,  .949} ${T_{\text DC}}$, $T_{\text BS}$ & 3 years, 10 years \\
		${Q_{{\text{th}}}}$, $a$, $b$, $E_{\text th}$ & 7, 2, 40, 1600 \\
		\rowcolor[rgb]{ .949,  .949,  .949} $ t_{\text infer}^{{\text{th}}} $, $ t_{\text trans}^{{\text{th}}} $, $ P_{{\text{trans}}}^{\max }$ & 0.3 s, 0.5 ms, 60 W \bigstrut[b]\\
		\hline
	\end{tabular}}%
	\label{tablexxx}%
\end{table}%

\subsubsection{Neural Network Structure}
The SNN-based actor network consists of two hidden layers, each with 256 spiking neurons \cite{chen2024fully}, where the spike information iterates over $ T_{\text{SNN}} =10 $ spike timesteps. The encoder module has $ N = 5 $ input populations, determined by the dimension of the state $ \boldsymbol{s} $ \cite{tang2021deep}. Each input  population contains $ N_{\text{e}} = 20 $ spiking neurons, resulting in $ N \times N_{\text{e}} $ spiking neurons in the input layer. The decoder module has $ M = 2 $ output populations, corresponding to the dimension of the action $ \boldsymbol{a} $ \cite{tang2021deep}. Each output population contains $ M_{\text{d}} = 10 $ spiking neurons, leading to $ M \times M_{\text{d}} $ spiking neurons in the output layer.
The double-critic network employs two independent DNNs, each consisting of two fully connected hidden layers with 256 neurons per layer and Mish activations. 
Both the actor and critic networks are trained using the Adam optimizer with learning rates $  \lambda_a = 1 \times 10^{-3} $ for the actor and $ \lambda_c = 1 \times 10^{-3} $ for the critic \cite{Kingma2014AdamAM}. The discount factor is set to $ \delta = 0.99 $, and the soft update rate for the target networks is $ \tau = 0.005 $. The replay buffer size is $ \left| \varepsilon \right| = 1 \times 10^{6} $, with a batch size of $ \left| {\cal B} \right| = 512 $. Exploration noise is introduced with a standard deviation of $ \sigma_0 = 0.1 $.

\subsubsection{Benchmark Soultions}
To demonstrate the effectiveness of the proposed SDRL algorithm, we have relied on three benchmark solutions:
\begin{itemize}
	\item \textit{Proximal Policy Optimization (PPO):} PPO is a widely used DRL algorithm that combines policy gradient methods with a clipped surrogate objective \cite{schulman2017proximal}. It is designed to ensure a stable and efficient training, making it a strong baseline for performance comparison.
	
	\item \textit{Soft Actor-Critic (SAC):} SAC is a state-of-the-art DRL algorithm that utilizes a stochastic policy and entropy-maximization objective \cite{haarnoja2018soft}. SAC's ability to balance exploration and exploitation makes it an effective benchmark for assessing SDRL’s performance in handling complex and dynamic environments.
	
	\item \textit{Random Policy:} This policy serves as a lower-bound baseline. Actions are selected randomly without considering the current state or optimization objectives, offering a contrast to learning-based solutions.	
\end{itemize}

\subsection{Results and Analysis}

\begin{figure}[t]
	\centering
	\subfigure[\label{FIG:compare-1}{Reward curves.}]{\includegraphics[width=0.31\textwidth]{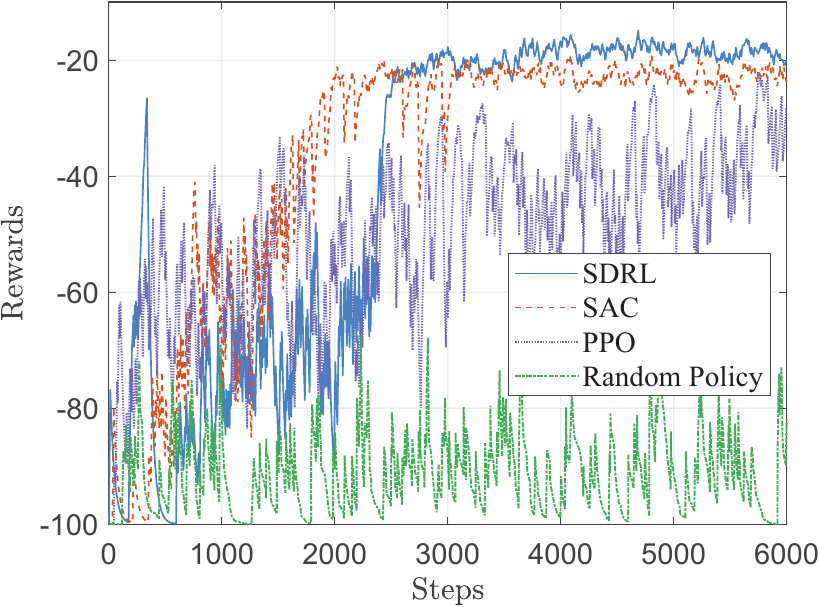}}
	\hspace{0.01\textwidth}
	\subfigure[\label{FIG:compare-2}{Average overall carbon footprint.}]{\includegraphics[width=0.15\textwidth]{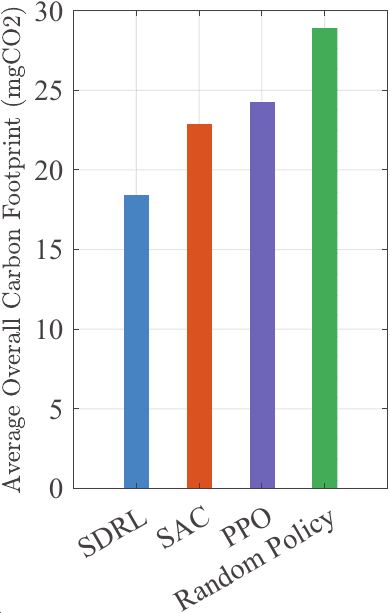}}
	\caption{Comparison of SDRL, SAC, PPO and random policy. The reward curves are smoothed for clarity.}
	\label{FIG:compare}
\end{figure}

Figure \ref{FIG:compare} illustrates the convergence behavior of the SDRL algorithm in comparison to the benchmarks SAC, PPO, and the random policy.
As shown in Fig. \ref{FIG:compare} (a), the proposed SDRL algorithm achieves the highest average reward (indicating the lowest overall carbon footprint) compared to the three benchmarks, with a convergence rate slightly slower than SAC.
Additionally, Figure \ref{FIG:compare} (b) reveals that the average overall carbon footprint achieved by the SDRL algorithm is reduced by 19.42\%, 24.08\%  and 36.30\% compared to SAC, PPO  and the random policy, respectively.
This demonstrates the significant benefits of the SNN-based actor in enhancing performance.
It is worth noting that results failing to meet the constraints in \eqref{problemCarbon} are excluded from the calculation of the average overall carbon footprint for PPO and the random policy, which further highlights the superior performance of the SDRL algorithm.
This improvement is largely attributed to the SNN’s event-driven computation, which reduces redundant calculations and optimizes computing resource usage, enabling superior policy optimization.
Furthermore, SNN’s ability to effectively handle spatiotemporal information allows the algorithm to better capture temporal dependencies within the environment, improving decision-making accuracy and contributing to the overall improved performance of the SDRL algorithm.

\begin{table}[t]
	\centering
	\caption{The average episodic return comparison between non-outage and outage strategies.}
	\small\begin{tabular}{ccc}
				\Xhline{0.65pt}
				Environments & SDRL  & SAC  \bigstrut\\
				\hline
				Outage Strategy & -63.19 & -50.30  \bigstrut\\
				Non-outage Strategy & -85.91 & -57.17  \\
				\Xhline{0.65pt}
	\end{tabular}%
	\label{tab:average_episodic_return}%
\end{table}

To compare the performance of non-outage and outage strategies, we present the average episodic returns of SDRL and SAC in Table \ref{tab:average_episodic_return}.
The average episodic return in this paper is defined as the average reward over the first 3000 steps, providing an indicator of early-stage learning efficiency \cite{yang2021wcsac}.
The carbon footprint model with non-outage strategy is derived from the model with outage strategy by setting $ \epsilon =0 $ in \eqref{eq_Cc_trans_15}. 
As shown in Table \ref{tab:average_episodic_return}, the outage strategy results in higher average episodic returns than the non-outage strategy, which indicates  that incorporating the outage strategy accelerates convergence, thereby reducing energy consumption and carbon emissions during the algorithm deployment. 
The performance improvement stems from the outage strategy’s ability to prevent excessive transmission delay and carbon emissions when the instantaneous SNR approaches zero, as demonstrated in \eqref{CarbonCommunTime} and \eqref{eq_Carbon_trans_1}.

\begin{figure*}[t]
	\centering
	\subfigure[\label{FIG:hiddenSize-1}Hidden layer size impact on reward curve.]
	{\includegraphics[width=0.325\textwidth]{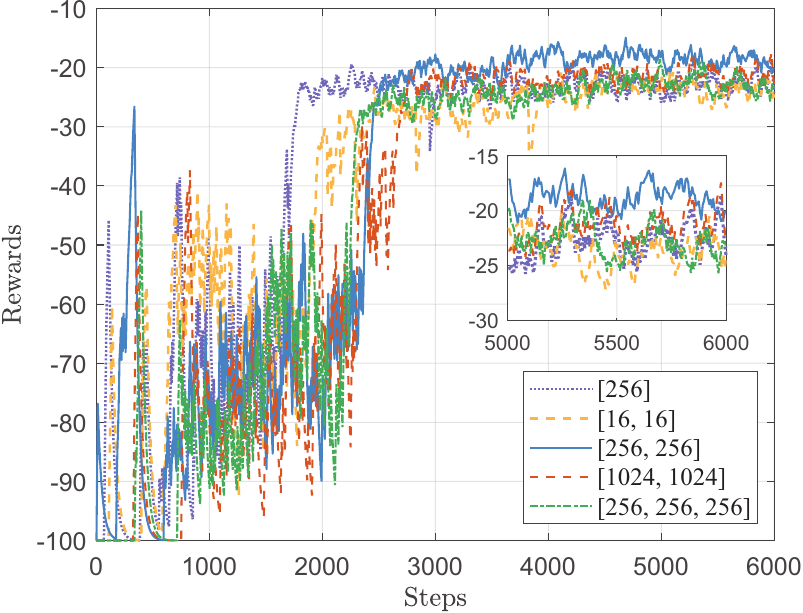}}
\hspace{0.02\textwidth}
	\subfigure[\label{FIG:hiddenSize-2and3}Hidden layer size impact on average overall carbon footprint.]
	{
		\begin{minipage}[b]{0.2\textwidth}
			\centering
			\includegraphics[width=\textwidth]{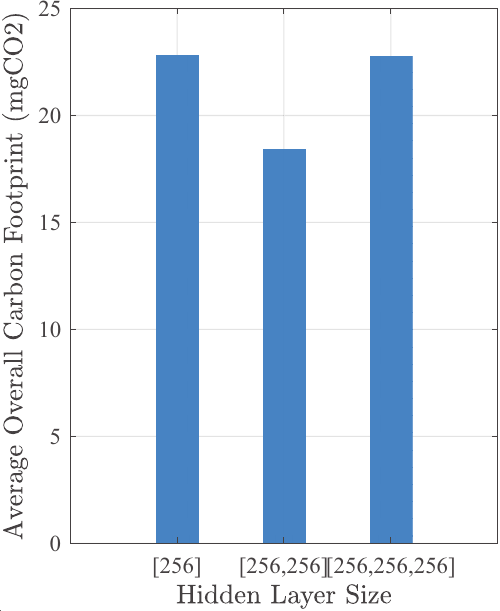}
		\end{minipage}
		\hspace{0.01\textwidth}
		\begin{minipage}[b]{0.38\textwidth}
			\centering
			\includegraphics[width=\textwidth]{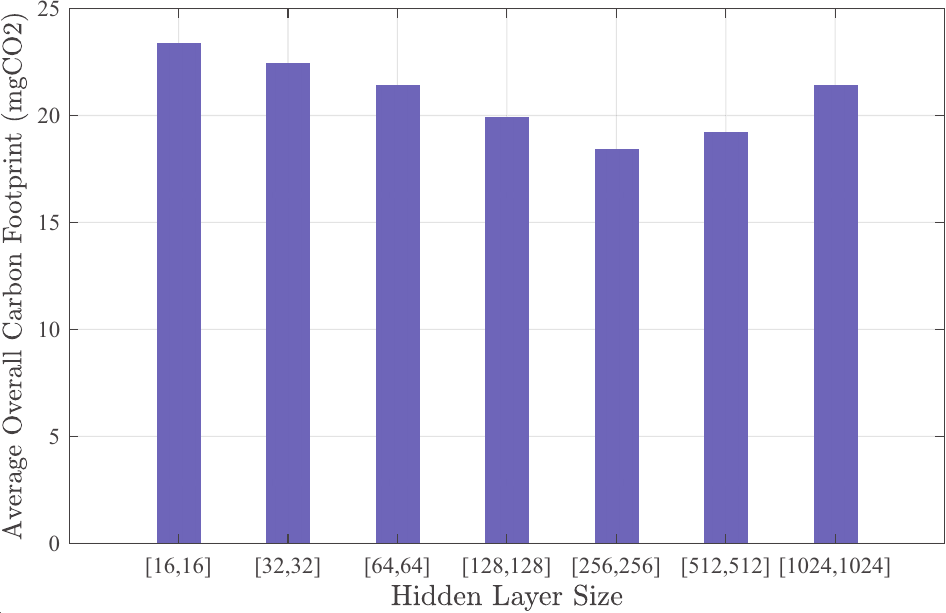}
		\end{minipage}
	}
	\caption{Impact of hidden layer size on reward curve and average overall carbon footprint. The reward curves are smoothed for clarity.}
	\label{FIG:hiddenSize}
\end{figure*}

Figure \ref{FIG:hiddenSize} illustrates the impact of the hidden layer size in the SNN-based actor of SDRL.
The optimal performance, in terms of average reward and overall carbon footprint, is achieved when the network contains two hidden layers, each with 256 spiking neurons.
The reason is that the initial increases in hidden layers and spiking neurons improve the ability to capture complex dependencies and extract features. However, further increases, beyond two hidden layers with 256 spiking neurons, lead to diminishing average rewards, which is likely due to overfitting or unnecessary computation without corresponding performance improvements.
When the number of spiking neurons in each hidden layer is reduced from 256 to 16, the average overall carbon footprint increases from 18.43 mgCO2 to 23.41 mgCO2, a rise of approximately 27.02\%. Despite this, the algorithm still converges successfully and provides a reasonable solution. This demonstrates that the SDRL algorithm is robust and capable of effectively adapting to smaller network sizes.

\begin{figure}[t]
	\centering
	\subfigure[\label{FIG:spike_ts}{Impact of spike timesteps on average reward and training time, whcih is normalized to its maximum value.}]{\includegraphics[width=0.247\textwidth]{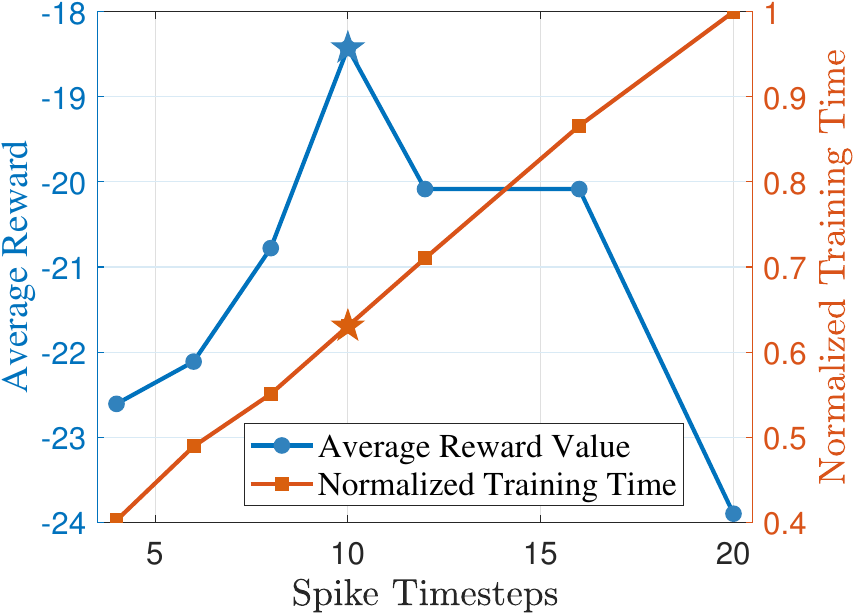}}
	\hspace{0.000001\textwidth}
	\subfigure[\label{FIG:popdim}
	{Impact of encoder and decoder population dimension on average reward.}]{\includegraphics[width=0.228\textwidth]{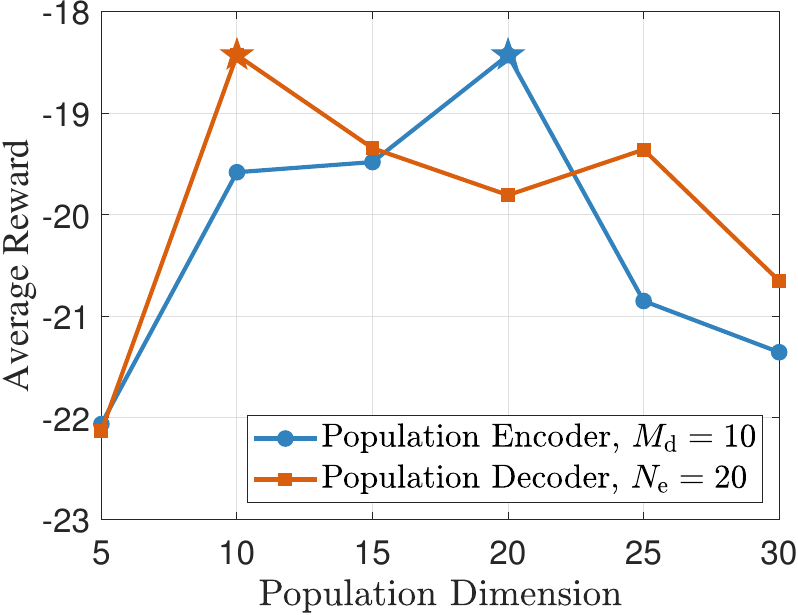}}
	\caption{Impact of spike timesteps and population dimension on convergence performance of SDRL.}
	\label{fig:7and8}
\end{figure}

The impact of spike timestep on the average reward and training time is illustrated in Fig. \ref{fig:7and8} (a). It can be observed that the average reward initially increases and then decreases as the spike timestep increases, indicating an optimal spike timestep, which appears to be around $ T_{\text{SNN}}=10  $. 
The normalized training time shows a steady increase as the spike timestep increases. 
When the spike timestep is too short, the model may lack sufficient temporal resolution, leading to a drop in performance. Conversely, excessively long timesteps may introduce unnecessary complexity, resulting in slower training without a corresponding performance improvement. 
This demonstrates SDRL's inherent adaptability: the spike timestep acts as a tunable factor to balance temporal resolution (critical for sequential decision-making) against computational overhead.
The observed optimal spike timestep $ T_{\text{SNN}}=10  $ reflects the balance between capturing enough temporal details and avoiding excessive complexity.
When extending the SDRL algorithm to other optimization problems, 
we recommend initializing SNN within this range and fine-tuning through multi-objective tradeoff analysis to balance reward gains against training time costs.

The impact of the population dimension on average reward for both encoder and decoder is depicted in Fig. \ref{fig:7and8} (b). The impact of the encoder population dimension is presented with a fixed decoder population dimension $M_{\text{d}} = 10$, while the impact of the decoder population dimension is demonstrated with a fixed encoder population dimension $N_{\text{e}} = 20$.
We observe that the average rewards initially increase and then decrease as the encoder and decoder population dimensions increase. 
The optimal average reward is achieved when the encoder population dimension is $N_{\text{e}} = 20$ and the decoder population dimension is $M_{\text{d}} = 10$.
The population dimensions control the number of spiking neurons in each input population of the encoder and each output population of the decoder.
The initial rise in average reward with increasing population dimensions is attributed to  the increased representational capability of the input and output spiking neuron populations. This allows for a more precise representation of states and actions, enabling the SNN-based actor to better capture sophisticated data dependencies.
However, beyond a certain point, further increases could introduce overfitting risks, where the network might capture noise rather than generalizable patterns. 
When applying the SDRL algorithm to other optimization problems, researchers should carefully determine population dimensions for both the encoder and decoder. 
The chosen population dimensions directly affect the state representation capacity and the action generalization ability of the SNN-based actor network. 
This parameter sensitivity highlights the importance of systematic hyperparameter tuning when deploying SDRL in new scenarios, which can be guided by our observed tradeoff between model ability for representation and generalization and overfitting risk.

\section{Conclusion}\label{sec:conclusion}
We proposed AOLO, a framework for analysis and  optimization for low-carbon oriented wireless LLM services, which introduces a comprehensive carbon footprint model incorporating greenhouse gas emissions from both the inference computation and wireless communication phases.
To provide more sustainable LLM inference services, 
we formulated an optimization problem aimed at  minimizing the overall carbon footprint, and proposed the SDRL algorithm, which leverages SNN as the actor.
Simulations demonstrate that SDRL significantly reduces the overall carbon footprint compared with the benchmark solutions.
Future work could further focus on low-carbon-oriented scheduling and resource allocation optimization in scenarios involving multiple LLM inference service providers and users, based on the comprehensive carbon footprint model proposed in this paper.

\begin{appendices}
\section{Proof of theorem \ref{The_averageT}}\label{Appendix_average}
\renewcommand{\theequation}{A-\arabic{equation}}
\setcounter{equation}{0}
Substituting \eqref{pdf_snr} and \eqref{CarbonCommunTime} into \eqref{eq_t_trans_14}, and substituting \eqref{pdf_snr} and \eqref{eq_Carbon_trans_1} into \eqref{eq_Cc_trans_15}, we have 
{\small\begin{align}
	&{{\bar t}_{{\text{trans}}}}  =\frac{1}{{1 -\epsilon }} {\left( {\frac{{m{\sigma ^2}}}{{\Omega P_{{\text{trans}}}}}} \right)^{\!\!m}}\!\frac{{\kappa \beta }}{{\Gamma \left( m \right)B}}{I_c},\\
	&{{\bar {\cal C}}_{\text{C}}}\! =\!\frac{1}{{1 \!-\!\epsilon }} \frac{{\kappa \beta \zeta_2 {m^m}{\sigma^{2m}}}}{{\Gamma \!\left( m \right){\Omega ^m}P_{{\text{trans}}}^{m - 1}B}}{I_c}\! +\! \frac{{\kappa \beta \!\left( {\zeta_2 {P_{{\text{fixed}}}}{T_{{\text{BS}}}} \!+\! {{\cal C}^{\text{E}}_{\text{BS}}}} \right)}}{{(1 -\epsilon )\mathcal{K}{T_{{\text{BS}}}}}}, 
\end{align}}
where 
{\small\begin{align}\label{eq_Ic_A3}
{I_c} = \int_{{\gamma _{{\text{th}}}}}^\infty  {\frac{{{\gamma ^{m - 1}}}}{{{{\log }_2}\left( {1 + \gamma } \right)}}\exp \left( { - \frac{{m{\sigma ^2}}}{{\Omega {P_{{\text{trans}}}}}}\gamma } \right)d\gamma }. \end{align}}
With the help of \cite[Eq.(8.4.6.5)]{prudnikov1990integrals} and \cite[Eq.(9.301)]{gradshteyn2007}, we have
{\small\begin{align}\label{eq_ln_zhankai}
\ln \left( {1 + \gamma } \right)  = \frac{1}{{2\pi i}}\int_{{{\cal R}_1}} {\frac{{\Gamma \left( {1 - {r_1}} \right){\Gamma ^2}\left( {{r_1}} \right)}}{{\Gamma \left( {1 + {r_1}} \right)}}{\gamma ^{{r_1}}}d{r_1}}. 
\end{align}}
Substituting \eqref{eq_ln_zhankai} into \eqref{eq_Ic_A3}, and with the help of \cite[01.03.07.0001.01]{webWolfram}, we can derive $ I_c $ as
{\small\begin{align}\label{eq_Ic_A5}
&{I_c}= \ln \!\left( 2 \right)\!\int_{{{\cal R}_1}} \!{\int_{{{\cal R}_2}} \!{\frac{{\Gamma \left( {1 + {r_1}} \right)}}{{\Gamma \left( {1 - {r_1}} \right){\Gamma ^2}\left( {{r_1}} \right)}}\Gamma \left( {{r_2}} \right)} } {\left( {\frac{{m{\sigma ^2}}}{{\Omega {P_{{\text{trans}}}}}}} \right)^{ - {r_2}}}\notag\\
&\qquad\qquad\qquad\qquad\times \underbrace{\int_{{\gamma _{{\text{th}}}}}^\infty  {{\gamma ^{m - 1 - {r_1} - {r_2}}}d\gamma }}_{\mathcal{I}_a} d{r_1}d{r_2},
\end{align}}
where $ {\mathcal{I}_a} = \frac{{\gamma _{{\rm{th}}}^{m - {r_1} - {r_2}}}}{{{r_1} + {r_2} - m}} = \frac{{\Gamma \left( {{r_1} + {r_2} - m} \right)}}{{\Gamma \left( {{r_1} + {r_2} - m + 1} \right)}}\gamma _{{\rm{th}}}^{m - {r_1} - {r_2}}, {r_1} + {r_2} - m > 0 $ is easily solved with the help of \cite[Eq.(8.331.1)]{gradshteyn2007}. 
Substituting $ {\mathcal{I}_a} $ into \eqref{eq_Ic_A5}, and with the help of the extended generalized bivariate Meijer G-function (EGBMGF) \cite[07.34.21.0081.01]{webWolfram}, we derive  the closed-form expression of $ {I_c} $ as \eqref{eq_Ic}, and then obtain the closed-form average transmission time and average carbon footprint of the wireless communication phase as \eqref{eq_t_trans_16} and \eqref{eq_Cc_17} respectively.
\end{appendices}

	\bibliographystyle{IEEEtran}
	\bibliography{CarbonRef}
	
\end{document}